\documentclass[twocolumn,amssymb, amsmath,nobibnotes,nofootinbib, aps, prl,showpacs]{revtex4}
\usepackage[dvips]{graphicx}
\begin{document}
\title{Exchange bias effect in the phase separated Nd$_{1-x}$Sr$_{x}$CoO$_3$ at the spontaneous ferromagnetic/ferrimagnetic interface}
\author{M. Patra, M. Thakur, S. Majumdar and S. Giri}
\email{sspsg2@iacs.res.in} 
\affiliation{Department of Solid State Physics and Center for Advanced Materials,Indian Association for the Cultivation of Science, Jadavpur, Kolkata 700 032, INDIA}

\begin{abstract}
We report the new results of exchange bias effect in Nd$_{1-x}$Sr$_{x}$CoO$_3$ for $x$ = 0.20 and 0.40, where the exchange bias phenomenon is involved with the ferrimagnetic (FI) state in a  spontaneously phase separated system. The zero-field cooled magnetization exhibits the FI ($T_{FI}$) and ferromagnetic ($T_C$) transitions at $\sim$ 23 and $\sim$ 70 K, respectively for $x$ = 0.20. The negative horizontal and positive vertical shifts of the magnetic hysteresis loops are observed when the system is cooled through $T_{FI}$ in presence of a positive static magnetic field. Training effect is observed for $x$ = 0.20, which could be interpreted by a spin configurational relaxation model. The unidirectional shifts of the hysteresis loops as a function of temperature exhibit the absence of exchange bias above $T_{FI}$ for $x$ = 0.20. The analysis of the cooling field dependence of exchange bias field and magnetization indicates that the ferromagnetic (FM) clusters consist of single magnetic domain with average size around $\sim$ 20 and $\sim$ 40 \AA ~ for $x$ = 0.20 and 0.40, respectively. The sizes of the FM clusters are close to the percolation threshold for $x$ = 0.20, which grow and coalesce to form the bigger size for $x$ = 0.40 resulting in  a weak exchange bias effect. 
                         
\end{abstract}
\pacs{75.70,Cn, 75.60.-d, 75.50.Gg}
\maketitle

\section{Introduction}
The exchange bias (EB) coupling in a heterogeneous system gives rise to the unidirectional anisotropy at the interface when the samples are cooled down to a temperature below a  critical temperature in an external magnetic field. \cite{meik} It manifests itself by a shift of the hysteresis loop and enhancement of coercivity, which attract considerable attention for the potential applications in magnetic memories, spin-electronics, and development of permanent magnet. Numerous reports are found on EB phenomenon at artificial interfaces, which have been mainly focused to develop advanced materials for the applications and understanding of the complex EB phenomenon. \cite{nogues1,kiwi,stamps,berk} Nevertheless, the signature of EB phenomenon is rarely observed in compounds with a unique crystal structure and having spontaneous interface. Long back the first example of EB effect without artificial interface was reported for Cu(Mn) and Ag(Mn) alloys, which have been recognized as typical spin-glass (SG) and cluster-glass (CG) systems depending on the dilution limit. \cite{kou} Recently, the signature of EB effect has been reported at the spontaneous interfaces for few mixed-valent manganites and cobaltites with perovskite structure. \cite{nieb,qian,patra,tang1,tang2,luo} The first report was found in a charge ordered (CO) compound  Pr$_{1/3}$Ca$_{2/3}$MnO$_3$,  where ferromagnetic (FM) droplets were spontaneously embedded in an antiferromagnetic (AFM)  background creating the  FM/AFM interface. \cite{nieb} The EB phenomenon has also been reported for another CO manganite Y$_{0.2}$Ca$_{0.8}$MnO$_3$, where a strong cooling field dependence of EB is observed due to a considerable change of phase fraction between FM/AFM layers. \cite{qian} Recently, we observed the signature of EB phenomenon in the cluster-glass (CG) compounds  LaMn$_{0.7}$Fe$_{0.3}$O$_3$ and La$_{0.87}$Mn$_{0.7}$Fe$_{0.3}$O$_3$, \cite{patra,kalyan} where short range FM clusters are embedded in a SG-like matrix creating spontaneous FM/SG interface. \cite{patra,de1,de2,liu} The average size of the FM clusters is found to decrease systematically with decreasing particle size, which has a strong  influence on the EB phenomenon. The EB field increases significantly with decreasing particle size, which was attributed to the increase of interface area by decreasing particle size. \cite{thakur2} The exchange bias-like phenomenon was also reported for another CG compound, La$_{0.80}$Ba$_{0.20}$CoO$_3$, which was suggested due to the freezing effect of the local anisotropy. \cite{luo} 

%1 one 2 two 3 three 4 four 5 five 6 six 7 seven 8 eight 9 nine 10 ten

The mixed-valent cobaltites with perovskite structure experience the delicate interplay among charge, spin state, transport, magnetic, and structural degrees of freedom exhibiting the complex phase separation scenario. An interesting phase diagram has been proposed for the hole doped compound Nd$_{1-x}$Sr$_x$CoO$_3$ depending on the degree of hole doping. \cite{stau} For low doping range (0 $< x <$ 0.18) the SG or CG state has been proposed with semiconducting temperature dependence of resistivity. With further increase in hole doping the short range FM clusters begin to coalesce above a percolation threshold ($x >$ 0.18)   to attain the magnetic long range ordering and start to show metallic conductivity in the ordered state. Metallic conductivity in both the paramagnetic and ordered states is observed for $x \geq$ 0.28. The coexistence of ferrimagnetic (FI) and FM ordering is reported for 0.20 $\leq x \leq$ 0.60. Neutron powder diffraction studies on Nd$_{0.67}$Sr$_{0.33}$CoO$_3$ confirm that the FM ($T_c$) and FI ($T_{\rm FI}$) ordering temperatures are $\sim$ 200 and $\sim$ 40 K, respectively, where ferrimagnetism was interpreted in terms of an induced antiparallel ordering of the Nd spins in close proximity of Co sublattice. \cite{krim} Several reports on electrical, magnetic, and thermodynamic studies of  Nd$_{1-x}$Sr$_x$CoO$_3$ also suggest the coexistence of FI and FM ordering for  $x$ = 0.33 and 0.50. \cite{para,yoshi,fonda} $^{59}$Co NMR studies on Nd$_{1-x}$Sr$_x$CoO$_3$ (0 $\leq x \leq$ 0.50) confirm different spin states of Co$^{3+}$ and Co$^{4+}$ ions. \cite{ghoshray} The parent compound NdCoO$_3$ shows the low spin (LS) state of Co$^{3+}$ ion in the paramagnetic state. As a result of hole doping (0.10 $\leq x \leq$ 0.20) an intermediate spin (IS) state of Co$^{3+}$ and Co$^{4+}$ appears in addition to the LS state of Co$^{3+}$  and the LS state of Co$^{3+}$ no more exists with the further increase in hole doping (0.30 $\leq x \leq$ 0.50). The reported results indicate that the magnetic and electronic phase separation scenario of Nd$_{1-x}$Sr$_x$CoO$_3$ is very similar to that in La$_{1-x}$Sr$_x$CoO$_3$, where the dissimilarity is that here Nd ion carries moment unlike La ion. \cite{wu} 

Recently, the signature of EB phenomenon ascribed to the intrinsic inhomogeneous phase separation was reported in the CG compounds La$_{1-x}$Sr$_x$CoO$_3$ (0.12 $\leq x \leq$ 0.30), where EB was suggested due to the cluster-glass state consisting of FM and SG phases. \cite{tang1,tang2} 
In order to observe EB phenomenon the system must involve with two exchange coupled phases, the reversible and rigid phases, where magnetization of the first one can be reversed and the second one can not be. The EB is observed for La$_{1-x}$Sr$_x$CoO$_3$ at the reversible FM and rigid SG interface. The phase separated compounds Nd$_{1-x}$Sr$_x$CoO$_3$ also exhibit the necessary ingredient for the EB effect, where spontaneous phase separation between FM and FI states exists giving rise to the FM/FI interface for $x \geq$ 0.18. Here, the EB phenomenon is investigated in Nd$_{1-x}$Sr$_x$CoO$_3$ with $x$ = 0.20 and 0.40, where the first one is close to the percolation threshold with semiconducting behavior and the second one exhibits the metallic conductivity. We observe the EB effect for both the samples, where the effect is strong for $x$ = 0.20 and weak for $x$ = 0.40. In order to explain different EB effects for both the compounds the nanoscale phase separation scenario has been proposed in Nd$_{1-x}$Sr$_x$CoO$_3$ for $x \geq$ 0.20. So far EB phenomenon has been reported for manganites and cobaltites with perovskite structures at spontaneous FM/AFM and FM/SG interface. Here, we present a new example of EB effect in the spontaneously phase separated compounds  Nd$_{1-x}$Sr$_x$CoO$_3$ for $x \geq$ 0.20, where EB effect is involved with ferrimagnetic states.

\begin{figure}[t]
%\vskip 0.4 cm
\centering
\includegraphics[width = 7 cm]{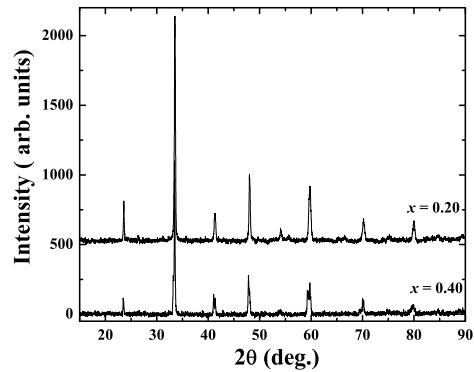}
\caption {X-ray powder diffraction patterns at room temperature for Nd$_{1-x}$Sr$_x$CoO$_3$ with $x$ = 0.20 and 0.40.}
\label{Fig. 1}
\end{figure}

\section{Experimental}

The polycrystalline samples of Nd$_{1-x}$Sr$_x$CoO$_3$ with $x$ = 0.20 and 0.40 were prepared by the chemical citrate route, which is described in our earlier report. \cite{de1} The stoichiometric proportion of Nd$_2$O$_3$, SrCO$_3$, and Co powders were dissolved in an  aqueous solution of nitric acid, mixed thoroughly, and then citric acid was added to achieve a homogeneous mixture of the metal citrates. The metal citrates were dried and decomposed at 873 K for 6 h. Finally, the powdered samples were pressed into  pellets and heated at 1273 K for 6 h followed by slow cooling at a rate of 0.7 K/min. The single orthorhombic ($Pbnm$) structure at room temperature is confirmed for both the cases, where x-ray powder diffraction (Seifert XRD 3000P) was recorded using CuK$_{\alpha}$ radiation. We do not observe any impurity phase in the x-ray diffraction patterns, which are shown in Fig. 1. The values of lattice parameters are 5.35 ($a$), 5.36 ($b$), and 7.56 ($c$) \AA ~for $x$ = 0.20, while for $x$ = 0.40 the parameters are 5.40 ($a$), 5.35 ($b$), and 7.61 ($c$) \AA.  The lattice parameters are consistent with the earlier report. \cite{ghoshray} The average size of the particles is found around $\sim$ 150.0 nm for both the cases, which was observed by a Transmission Electron Microscope (TEM), model ZEOL JEM-2010. The dc magnetization was measured using a commercial superconducting quantum interference device (SQUID) magnetometer (MPMS, XL). The sample was cooled down to the lowest temperature in zero magnetic field and the magnetization was measured in the warming cycle by applying an external magnetic field for zero-field cooled (ZFC) magnetization. On the other hand, the sample was cooled in non-zero field for the measurement of field-cooled (FC) magnetization and the measurement was performed in the warming cycle with field kept switched on.  

\begin{figure}[t]
%\vskip 0.4 cm
\centering
\includegraphics[width = 9 cm]{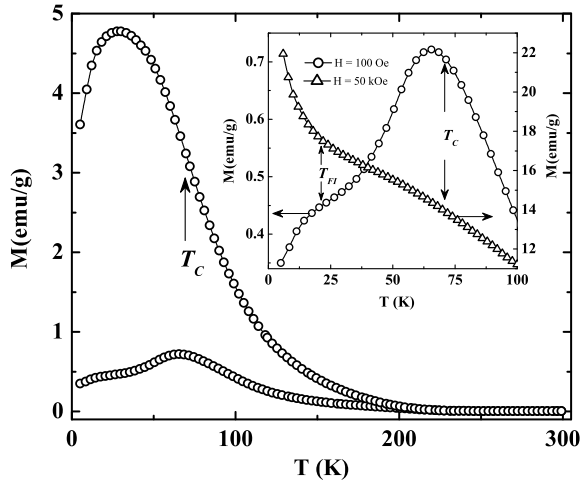}
\caption {Temperature dependence of magnetization measured at 100 Oe under zero field cooled (ZFC) and field cooled (FC) conditions for $x$ = 0.20. The inset shows the ZFC magnetization highlighting ferrimagnetic ($T_{FI}$) and ferromagnetic ($T_c$) transitions for the measurements at 100 Oe and 50 kOe.}
\label{Fig. 2}
\end{figure}

\begin{figure}[t]
%\vskip 0.4 cm
\centering
\includegraphics[width = 10 cm]{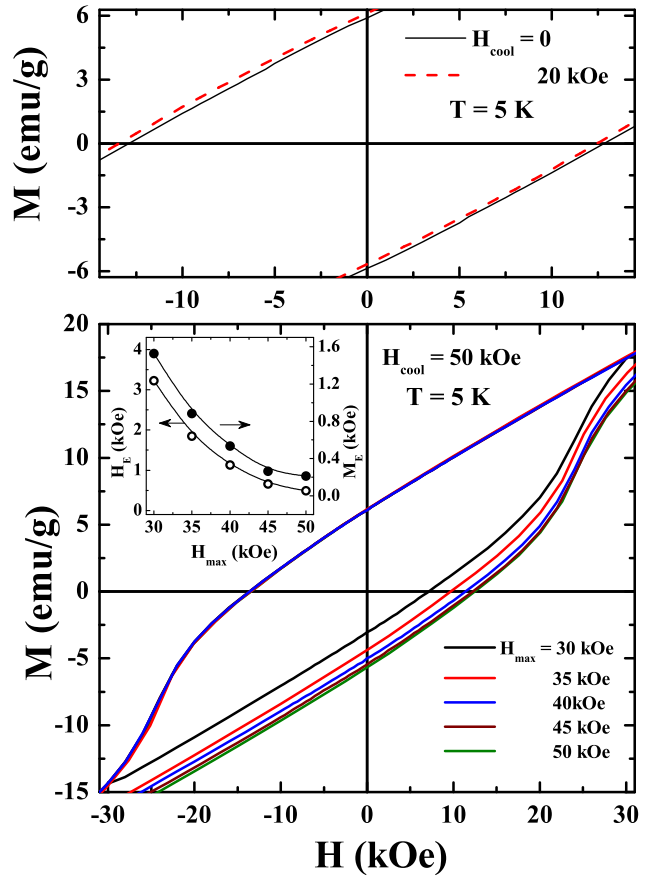}
\caption {(Color online) In the top panel central part of the hysteresis loops at 5 K  measured after cooling the sample in zero field and field at 20 kOe, where loop measured after field cooling are indicated by the broken curve. In the bottom panel central part of the minor loops at 5 K for the measurements in between $\pm$ 30, $\pm$ 35, $\pm$ 40, $\pm$ 45, and $\pm$ 50 kOe after cooling the sample with 50 kOe. The inset exhibits the variation of exchange bias field ($H_E$) and magnetization ($M_E$) with $H_{max}$ for $x$ = 0.20.}
\label{Fig. 3}
\end{figure}

\section{Experimental results}

Temperature dependence of field-cooled effect of magnetization measured at 100 Oe is shown in Fig. 2 for Nd$_{0.80}$Sr$_{0.20}$CoO$_3$. The inset of the figure exhibits ZFC magnetization highlighting the ferrimagnetic ($T_{FI}$) and ferromagnetic ($T_c$) transitions at $\sim$ 23 and $\sim$ 70 K, respectively, where $T_{FI}$ is indicated by a  shoulder around $\sim$ 23 K for the low-field measurement at 100 Oe. A sharp increase in the ZFC magnetization is observed below $T_{FI}$ when the measurement was performed at a high field with 50 kOe. $T_c$ is defined around $\sim$ 70 K, where a sharp increase in FC magnetization is observed for the measurement at 100 Oe. The values of $T_{FI}$ and $T_c$ in the present observation are reproduced exactly as compared to the values reported by Stauffer {\it et al}. \cite{stau} 
The reported experimental results including the neutron and $^{59}$Co NMR results clearly indicate the coexistence of FM and FI phases at low temperature in Nd$_{1-x}$Sr$_{x}$CoO$_3$ for $x \geq$ 0.20. \cite{stau,krim,ghoshray,para,yoshi,fonda} Thus, the magnetization loops at 5 K should have two components corresponding to the FM and FI phases. Recently, Niebieskikwiat and Salamon reported the EB effect in Pr$_{1/3}$Ca$_{2/3}$MnO$_3$ consisting of FM and AFM phases at low temperature, where the AFM background exhibiting linear $M-H$ response was subtracted from the overall magnetization loop at 5 K to get the evident features of the horizontal and vertical shifts. \cite{nieb} In the present investigation it is difficult to subtract the FI background. Nevertheless, the horizontal and vertical shifts of the hysteresis loop are clearly observed exhibiting the typical feature of EB effect when the sample was cooled in a non-zero field. For non-zero positive cooling field a negative shift in the field axis and a positive shift in the magnetization axis are observed at 5 K with respect to the hysteresis loop for $H_{cool}$ = 0, which is shown in the top panel of Fig. 3.   
 We define the coercivity, $H_C$ = $\left|(H_{\rm left} - H_{\rm right})/2\right|$
 $\approx$ 13 kOe and EB field, -$H_E$ = ($H_{\rm left}$ + $H_{\rm right}$)/2 $\approx$ - 500 Oe for the measurement of magnetic hysteresis in between $\pm$ 50 kOe. Here, $H_{\rm right}$ and $H_{\rm left}$ are the positive and negative values of field at magnetization, $M$ = 0, respectively. The remanence asymmetry is defined as EB magnetization ($M_E$), which is estimated from the vertical shifts at the saturation. \cite{nogues3} The value of $M_{E}/M_{S}$ is $\sim$ 0.53 $\times$ 10$^{-2}$, where $M_S$ is the saturation of  magnetization. The values of the asymmetry parameters associated with the FM and FI states are comparable to those involved with the FM and SG states for the cobaltite   La$_{0.88}$Sr$_{0.12}$CoO$_3$, where the maximum effect of EB was reported in the series.  \cite{tang1,tang2} The values of $H_E$ and $M_{E}/M_{S}$ were 500 Oe and 1.10 $\times$ 10$^{-2}$, respectively when hysteresis was measured in between $\pm$ 50 kOe for  La$_{0.88}$Sr$_{0.12}$CoO$_3$. \cite{tang2} The reported values of the shifts for the phase separated cobaltites are compared in Table 1 with the present observations for $x$ = 0.20 and 0.40 for different $H_{max}$, where $H_{max}$ is the maximum field used for the measurement of hysteresis loops. 

\begin{table}
{\bf Table 1:} The maximum values of exchange bias field ($H_E$) and relative vertical shift ($M_{E}/M_{S}$) at different types of interfaces for phase separated cobaltites measured in between different $\pm$ $H_{max}$.

\begin{tabular}{lcccc}
\hline
\hline
%\begin{ruledtabular}
System                                    & $H_{max}$ & $H_E$ & $M_{E}/M_{S}$ & Interface  \\
  
                                          & kOe       & kOe    & (10$^{-2}$) &    \\
\hline
La$_{0.80}$Ba$_{0.20}$CoO$_3$\cite{luo}   & 3       & 0.54   & -   & FM/SG \\
La$_{0.88}$Sr$_{0.12}$CoO$_3$\cite{patra1} & 20     & 2.95   & 10.40   & FM/SG \\
La$_{0.88}$Sr$_{0.12}$CoO$_3$\cite{tang2}	 & 50     & 0.50	 & 1.10	  & FM/SG \\		Nd$_{0.80}$Sr$_{0.20}$CoO$_3$$^*$          & 30     & 3.23   & 4.15   & FM/SG \\
																					 & 50     & 0.50   & 0.53   & FM/FI \\
Nd$_{0.60}$Sr$_{0.40}$CoO$_3$$^*$          & 50     & 0.08   & 0.55   & FM/FI \\	 
\hline 
\hline 
\end{tabular}

$^*$ Current investigation
\end{table}

The characteristics of EB phenomenon strongly depend on the delicate interplay among Zeeman energy ($E_Z$) associated with the FM clusters, anisotropy energy ($E_A$)  associated with the FI clusters, and exchange energy ($E_{ex}$) at the FM/FI interface. The first term can be experimentally controlled by tuning $H_{cool}$ and $H_{max}$. The magnetic hysteresis loops were measured for $x$ = 0.20 at 5 K in between $\pm H_{max}$ = $\pm$ 30, $\pm$ 35, $\pm$ 40, $\pm$ 45, and $\pm$ 50 kOe when the sample was cooled down to 5 K from 250 K in presence of 50 kOe field. The central parts of the hysteresis loops are shown in the bottom panel of Fig. 3. The values of $H_E$ and $M_E$ as a function of $H_{max}$ are shown in the inset of the figure. The values of $H_E$ and $M_E$ are decreased with increasing $H_{max}$. The plots further indicate that the values show a tendency of stabilization for $H_{max} \geq$ 45 kOe. We note that the hysteresis loops measured in between $\pm$ 45 and $\pm$ 50 kOe almost merge  together. The $H_{max}$ dependence of $H_E$ and $M_E$ has been reported by Tang {\it et al}. for La$_{0.82}$Sr$_{0.18}$CoO$_3$, where $H_E$ and $M_E$ vanish even at small value of $H_{max}$ = 30 kOe. \cite{tang1} The results are analogous to the minor loops effect, which is explicitly described by Geshev. \cite{gesh} However, the signature of EB effect in La$_{0.82}$Sr$_{0.18}$CoO$_3$ was confirmed by the training effect. \cite{tang1} The $H_{max}$ dependence of $H_E$ and $M_E$ has also been investigated by Salazar-Alvarez {\it et al}. for MnO/Mn$_3$O$_4$ core/shell structure, where the plots show the non-zero asymptotic value of the horizontal shift while the vertical shifts virtually vanish at $H_{max} >$ 70 kOe. \cite{nogues}   
In the present observation the plots show that $H_E$ and $M_E$ stabilize around $H_{max}$ =  50 kOe, where the values of $H_E$ and $M_E$ are considerably large. The results indicate that the plots are not the typical minor loops effect described by Geshev \cite{gesh}, rather it exhibits significant EB effect.    

\begin{figure}[t]
%\vskip 0.4 cm
\centering
\includegraphics[width = 8 cm]{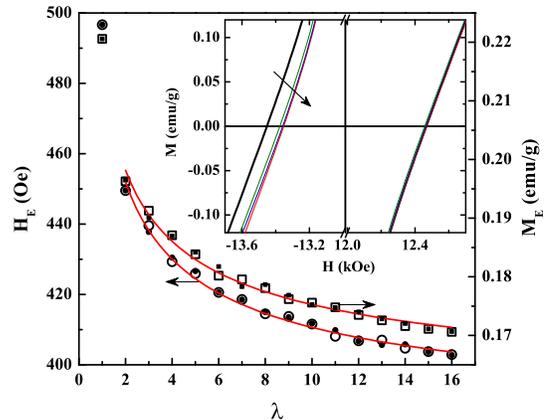}
\caption {(Color online) The decrease of exchange bias field ($H_E$) and magnetization ($M_E$) with the consecutive number ($\lambda$) of cycling of the hysteresis loop for $x$ = 0.20 exhibiting training effect. The central part of the 1st, 2nd, 8th, and 16th loops are shown in the inset, where arrow indicates the increasing direction of $\lambda$.}
\label{Fig. 4}
\end{figure}

Training effect is one of the important features of the exchange bias system, which describes the decrease of $H_C$, $H_E$, and $M_E$ when the system is successively field cycled at a particular temperature after the field cooling. \cite{nogues1,nieb,tang1} The training effect is investigated for $x$ = 0.20, where the sample was cooled down to 5 K from 250 K in presence of 20 kOe field and then the hysteresis loops were measured successively in between $\pm$ 50 kOe up to 16 times. We note that the training effect is evident in the system. The central parts of the 1st, 2nd, 8th, and 16th loops are shown in the inset of Fig. 4. The systematic decrease of $H_E$ and $M_E$ with the consecutive cycling number ($\lambda$) are shown in Fig. 4 by the open symbols. The values of $H_E$ and $M_E$ are decreased up to $\sim$ 10 \% and $\sim$ 11 \%, respectively for $\lambda$ = 2, which are comparable to the reported results for the spontaneously phase separated systems. For Pr$_{1/3}$Ca$_{2/3}$MnO$_3$ \cite{nieb} bearing the FM/AFM interface   the value of $M_E$ was decreased up to $\sim$ 20 \% for the second cycle, whereas the values of $M_E$ and $H_E$ were decreased up to $\sim$ 10 \% and $\sim$ 19 \%, respectively for La$_{0.82}$Sr$_{0.18}$CoO$_3$ having FM/SG interface. \cite{tang1}   

The decrease of $H_E$ ($M_E$) is fitted satisfactorily with the following empirical  relation 
\begin{equation}
H_E(\lambda) - H_E^{\infty} \propto \frac{1}{\sqrt{\lambda}},
\end{equation} 
where $H_E^{\infty}$ is the value for $\lambda \rightarrow \infty$. The solid lines in Fig. 4 exhibit the best fit of $H_E$ and $M_E$ variation with $\lambda$ for $\lambda \geq$ 2. The values of the fitted parameters are $H_E^{\infty} \approx$ 377 Oe and $M_E^{\infty} \approx$ 0.157 emu/g. The above empirical relation does not fit the sharp decrease between first and second loops, which is in accordance with the reported results. \cite{nogues1,nieb,tang1} Recently, Binek proposed a recursive formula in the framework of spin configurational relaxation to understand the training effect for a FM/AFM heterostructure, which describes the ($\lambda$+1)th loop shift with respect to the $\lambda$th one as \cite{binek}
\begin{equation}
H_E (\lambda + 1) - H_E (\lambda) = -\gamma [H_E (\lambda) - H_E^{\infty\prime}]^3,
\end{equation} 
where $\gamma$ is a sample dependent constant. Using $\gamma$ = 2.28 $\times$ 10$^{-5}$ Oe$^{-2}$ and $H_E^{\infty\prime}$ = 369.4 Oe the whole set of data is reproduced for $H_E$, while the values of $\gamma$ and $M_E^{\infty\prime}$ are 79.5 (emu/g)$^{-2}$ and 0.153 emu/g, respectively for $M_E$. The calculated values are shown by the filled symbols in Fig. 4, which matches satisfactorily with the experimental data (open symbols). Thus, the spin configurational relaxation model can describe our experimental results satisfactorily, where successive reversal of the FM spins triggers the configurational relaxation of the interfacial FI spins toward equilibrium giving rise to the training effect.

\begin{figure}[t]
%\vskip 0.4 cm
\centering
\includegraphics[width = 7 cm]{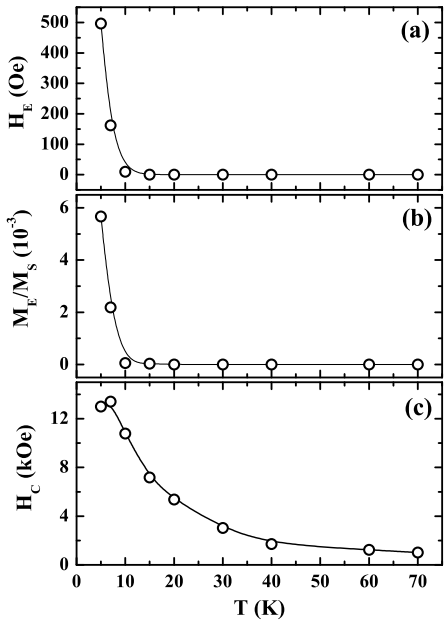}
\caption {Temperature dependence of (a) exchange bias field ($H_E$), (b) relative vertical shift ($M_{E}/M_{S}$), and (c) coercivity ($H_C$) with $H_{cool}$ = 20 kOe for $x$ = 0.20.}
\label{Fig. 5}
\end{figure}

The magnetic hysteresis loops were measured in between $\pm$ 50 kOe at different temperatures for Nd$_{0.80}$Sr$_{0.20}$CoO$_3$ with the sample been cooled down to the desired temperatures from 250 K with $H_{cool}$ = 20 kOe. The values of $H_C$, $H_E$, and $M_{E}/M_{S}$ as a function of temperature are shown in Fig. 5. $H_C$ decreases slowly and vanishes at the onset of the Curie temperature around $\sim$ 70 K. On the other hand, $H_E$ and $M_{E}/M_{S}$ decrease sharply with increasing temperature and disappear  for $T \geq$ 20 K, where a shoulder in the ZFC magnetization is observed indicating the signature of ferrimagnetic ordering. The temperature dependence of the asymmetry parameters  are typical for the exchange biased systems viz., charge ordered Pr$_{1/3}$Ca$_{2/3}$MnO$_3$, \cite{nieb} cluster-glass LaMn$_{0.7}$Fe$_{0.3}$O$_3$,  \cite{patra} and cluster-glass cobaltites La$_{1-x}$Sr$_x$CoO$_3$, \cite{tang1,tang2} where EB effect vanishes above AFM ($T_N$) or spin freezing ($T_f$) transition temperature. In the cases of EB systems the rigid AFM or SG spins apply a coupling  force on the FM spins at the interface and a layer of pinned or frozen FM spins are created on the outer surface of the FM clusters when the system is cooled in non-zero field. The pinned or frozen FM spins give rise to the unidirectional shift of the hysteresis loops and reveal the EB effect. For Nd$_{1-x}$Sr$_{x}$CoO$_3$ with $x \geq$ 0.20 the coexistence of FM and FI phases has been suggested by Stauffer {\it et al}., \cite{stau} which is also indicated here by the ZFC magnetization for $x$ = 0.20 (inset of Fig. 2).  Therefore, the  EB effect below $T_{FI}$ is suggested due to the pinning effect. The rigid FI spins operate the pinning force on the reversible FM spins at the FM/FI interface and the pinned FM spins lead to the EB effect. The signature of EB effects involving with the ferrimagnetic phase are reported at the different combinations of artificial  interfaces viz., FM/FI, FI/AFM, and FI/FI interface for different bilayer heterogeneous structures. \cite{nogues1,canet,mangin,berk1,hauet,nogues} Here, we observe the new results of EB effect at the intrinsic FM/FI interface, where the EB effect is ascribed to the spontaneous separation between FM and FI phases.   

\begin{figure}[t]
%\vskip 0.4 cm
\centering
\includegraphics[width = 8 cm]{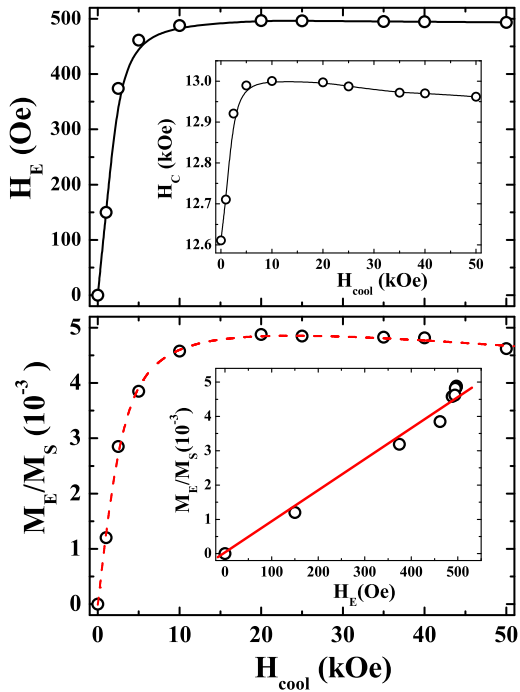}
\caption {(Color online) Cooling field ($H_{cool}$) dependence of (top panel) exchange bias field ($H_E$) and (bottom panel) $M_{E}/M_{S}$ at 5 K for $x$ = 0.20, where the hysteresis loops were measured between $\pm$ 50 kOe. $M_E$ and $M_S$ are the exchange bias magnetization and saturation magnetization. The $H_{cool}$ dependence of coercivity ($H_C$) and the plot of $M_{E}/M_{S}$ against $H_E$ are shown in the inset of top and bottom panels, respectively. The broken curve in the bottom panel is the fitted curve using simplified exchange interaction model. The solid straight line in the inset of the bottom panel shows the linear fit.}
\label{Fig. 6}
\end{figure}

\begin{figure}[t]
%\vskip 0.4 cm
\centering
\includegraphics[width = 8.5 cm]{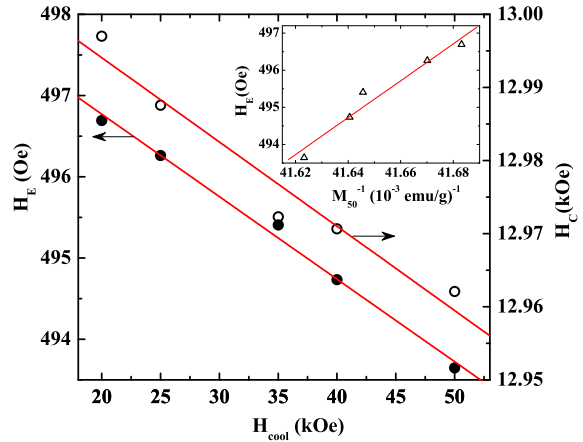}
\caption {(Color online) Plots of exchange bias field ($H_E$) and coercivity ($H_C$) against cooling field ($H_{cool}$) at 5 K for $x$ = 0.20. Inset shows the plot of $H_E$ as a function of inverse of the average magnetization at 50 kOe ($M_{50}^{-1}$). Solid straight lines indicate the linear fits.}
\label{Fig. 7}
\end{figure}

The cooling field dependence of EB effect was investigated for $x$ = 0.20, where the sample was cooled down to 5 K from 250 K with different cooling fields and the hysteresis loops were measured in between $\pm$ 50 kOe. The $H_{cool}$ dependence of $H_E$ is shown in the top panel of Fig. 6. The value of $H_E$ increases sharply with $H_{cool}$ up to 10 kOe and then shows a slight decreasing trend above 20 kOe. As seen in the inset of the top panel of the figure the $H_{cool}$ dependence of $H_C$ follows similar behavior like the $H_{cool}$ dependence of $H_E$, where the sharp increase of $H_C$ is associated with the increase of $H_E$. The plots of $H_E$ and $H_C$ with $H_{cool}$ exhibit that it follows nearly linear dependence above 20 kOe, which is shown in Fig. 7. Recently, Qian {\it et al}. have shown the linear $H_{cool}$ dependence of $H_E$ above 10 kOe, where strong correlation between $H_E$ and $H_C$ was reported. \cite{qian} The authors further suggested that the decrease of $H_E$ with increasing $H_{cool}$ is associated with the increase of the size of the FM layers  for Y$_{0.2}$Ca$_{0.8}$MnO$_3$. Considering the magnetization at 50 kOe close to the saturation value at 5 K, the increase of $M_{50}$  indicates the increase of the average size of the FM clusters, where $M_{50}$ is determined from the average value of the positive and negative values of the magnetization at 50 kOe. The linear plot of $H_E$ against $M_{50}^{-1}$ is shown in the inset of Fig. 7, which indicates that $H_E$ is inversely proportional to the size of the FM clusters. Note that a very small decrease of $H_E$ around $\sim$ 0.6 \% is observed at 5 K for the increase of $H_{cool}$ from 20 kOe to 50 kOe. 
In the high $H_{cool}$ range the feature of $H_E$ is reminiscent to that observed for FM/AFM bilayers, \cite{cai} AFM-core and FI-shell structure, \cite{nogues} and unlike to that observed for FM/SG  systems. \cite{patra,thakur2,tang1,tang2,bianco} For the exchange biased systems involved with a SG state the values of $H_E$ are decreased considerably with increasing $H_{cool}$ above a certain value of $H_{cool}$, where large $H_{cool}$ typically affects the frozen spins in the  SG phase and frozen FM spins by polarizing the glassy magnetic spins toward the direction of $H_{cool}$. In fact, $H_E$ decreases up to $\sim$ 40 \% for the increase of $H_{cool}$ from 20 to 50 kOe in La$_{0.88}$Sr$_{0.12}$CoO$_3$, \cite{tang2}  $\sim$ 47 \% for the increase of $H_{cool}$ from 6 to 12 kOe in  LaMn$_{0.7}$Fe$_{0.3}$O$_3$, \cite{patra} and $\sim$ 25 \% for the increase of $H_{cool}$ from 2 to 4 kOe in La$_{0.87}$Mn$_{0.7}$Fe$_{0.3}$O$_3$, \cite{kalyan} where the systems are involved with the SG states exhibiting EB effect. The frozen FM spins and pinned FM spins may have different features of cooling field dependence, where the large values of $H_{cool}$  can not influence the pinned FM spins significantly, which is indicated by the minor change of $H_E$ for AFM or FI state. The large reduction of $H_E$ around $\sim$ 37 \% is exceptionally observed in Y$_{0.2}$Ca$_{0.8}$MnO$_3$ for the increase of $H_{cool}$ from 10 kOe to 60 kOe, which is associated with the FM/AFM interface. In the above case the large decrease of $H_E$ was attributed to the increase of the phase fraction of the FM layer at the cost of AFM layer, which was pointed out by Qian {\it et al}. \cite{qian} Herein, the minor reduction of $H_E$ around $\sim$ 0.6 \% may indicate that small increase of the size of FM clusters is ascribed to the reduction of pinned FM layers at the FM/FI interface, which does not involve with the conversion of phase fraction. 

Recently, a simplified exchange interaction model has been proposed by Niebieskikwiat and  Salamon considering the FM clusters to be embedded in the AFM host for charge ordered compound   Pr$_{1/3}$Ca$_{2/3}$MnO$_3$. \cite{nieb} The FM clusters were assumed to have a single domain structure, which exhibit magnetization reversal like FM nanoparticles consisting of single magnetic domain. The model gives the simplified relation between $H_E$ and $M_E$/$M_S$ as $H_E$ $\propto$ $M_E$/$M_S$ for $\mu H_E <$ k$_{B}T$, where $M_S$ and $\mu$ are the saturation magnetization and average moment of the FM clusters, respectively. We observe nearly linear behavior between $H_E$ and $M_E$/$M_S$ in the inset of the bottom panel of Fig. 6, which has also been verified for Pr$_{1/3}$Ca$_{2/3}$MnO$_3$, \cite{nieb} CaMnO$_{3-\delta}$, \cite{mark} LaMn$_{0.7}$Fe$_{0.3}$O$_3$,  \cite{patra,thakur2} and  La$_{0.87}$Mn$_{0.7}$Fe$_{0.3}$O$_3$ \cite{kalyan} exhibiting the EB effect. Here, the value of $\mu H_E$/k$_{B}T$ $\approx$ 0.87, where the value of $\mu$ is obtained from the fit of $M_{E}/M_{S}$ against $H_{cool}$ plot, which is described below. Note that the value of $\mu H_E$/k$_{B}T$ was $\sim$ 0.9 for Pr$_{1/3}$Ca$_{2/3}$MnO$_3$. \cite{nieb} The linear dependence of the $M_E$/$M_S$ against $H_E$ plot indicates that FM clusters may have a single domain structure for Nd$_{0.80}$Sr$_{0.20}$CoO$_3$. The model further defines the $H_{cool}$ dependence of $M_{E}/M_{S}$ as 
\begin{equation}
H_E \propto M_{E}/M_{S} \propto j_i\left[\frac {j_i\mu_0}{(g \mu_B)^2}L(\frac { \mu H_{cool}}{k_BT_f})+H_{cool}\right], 
\end{equation}
which was verified successfully for spontaneously phase separated Pr$_{1/3}$Ca$_{2/3}$MnO$_3$, \cite{nieb}  LaMn$_{0.7}$Fe$_{0.3}$O$_3$, \cite{patra,thakur2} and  La$_{0.87}$Mn$_{0.7}$Fe$_{0.3}$O$_3$. \cite{kalyan} The first term in the expression dominates for small $H_{cool}$, whereas the second term dominates for large $H_{cool}$, which varies linearly with $H_{cool}$. In the above relation $J_i$ and $\mu$ are adjustable parameters,  where $J_i$ is the interface exchange constant. $T_f$ in the above expression was defined as  the freezing temperature below which $M_{E}/M_{S}$ was found to increase steeply for Pr$_{1/3}$Ca$_{2/3}$MnO$_3$. Here, we assume the value of $T_f$ $\sim$ 15 K below which $M_{E}/M_{S}$ increases sharply. The broken curve in the bottom panel of Fig. 6 exhibits a satisfactory fit of the experimental data using the above expression. The number density of FM clusters ($n$) is estimated from the saturation magnetization, $M_S \approx n\mu$, where the value of $M_S$ at 5 K is obtained from the extrapolation of the magnetization to 1/$H$ $\rightarrow$ 0. The value of $n$ is $\approx$ 15.5 $\times 10^{-5}$ \AA$^{-3}$ which further gives the rough estimate of the size of the FM clusters around $\sim$ 20 \AA. The value of the size of FM cluster is consistent with those around $\sim$ 10 \AA ~for Pr$_{1/3}$Ca$_{2/3}$MnO$_3$ \cite{nieb} and $\sim$ 10 - 30 \AA ~for LaMn$_{0.7}$Fe$_{0.3}$O$_3$ depending on the particle size. \cite{thakur2} 

\begin{figure}[t]
%\vskip 0.4 cm
\centering
\includegraphics[width = 8.5 cm]{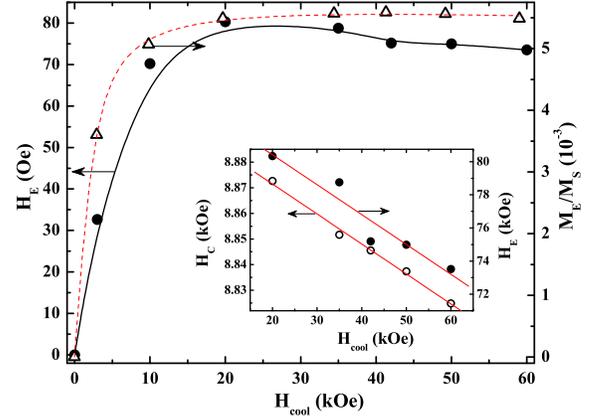}
\caption {(Color online) Plots of exchange bias field ($H_E$) and relative vertical shift ($M_{E}/M_{S}$) against cooling field ($H_{cool}$) at 5 K for $x$ = 0.40. The broken curve is the fit of the $H_{cool}$ dependence of $M_{E}/M_{S}$ plot. Inset shows the plot of $H_E$ and $H_C$ as a function of $H_{cool}$. Solid straight lines indicate the linear fits.}
\label{Fig. 8}
\end{figure}

We also investigate the EB effect for $x$ = 0.40, where the maximum values of $H_E$ and $M_{E}/M_{S}$ are shown in Table 1. The value of $M_{E}/M_{S}$ is nearly the same, whereas $H_E$ is much smaller for $x$ = 0.40 than that of the value for $x$ = 0.20. The $H_{cool}$ dependence of EB effect was also measured for $x$ = 0.40. The plots of $H_E$ and $M_{E}/M_{S}$ as a function of $H_{cool}$ at 5 K are shown in Fig. 8. $H_E$ is found to increase sharply up to $\sim$ 20 kOe and shows a decreasing trend with the further increase of $H_{cool}$, where $M_{E}/M_{S}$ follows almost similar behavior. The linear dependence of $M_{E}/M_{S}$ against $H_E$ is observed for $x$ = 0.40 analogous to $x$ = 0.20, which indicates that the FM clusters of $x$ = 0.40 also have a single domain structure. We observe the linear $H_{cool}$ dependence of $H_E$ and $H_C$ above 20 kOe, which is shown in the inset of Fig. 8. Here, a very small decrease of $H_E$ ($\sim$ 0.5 \%) is also observed for $x$ = 0.40 like $x$ = 0.20. The model proposed by Niebieskikwiat and Salamon \cite{nieb} was used to fit the $H_{cool}$ dependence of $M_{E}/M_{S}$, where the satisfactory fit is shown in Fig. 8 by the broken curve. The number density of FM clusters is thus obtained from the fit as $n$ $\approx$ 32.5  $\times 10^{-5}$ \AA$^{-3}$, which gives the rough estimate of the size of the FM clusters around $\sim$ 40 \AA. The analysis indicates that the average size of the FM clusters is increased considerably for $x$ = 0.40 than $x$ = 0.20.

\begin{figure}[t]
%\vskip 0.4 cm
\centering
\includegraphics[width = 8.5 cm]{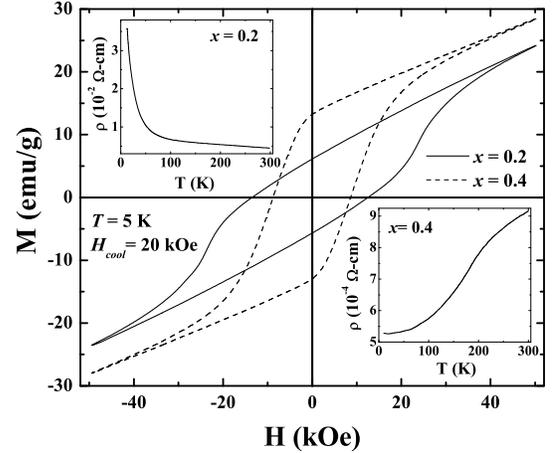}
\caption {Magnetic hysteresis loop at 5 K after field cooling with 20 kOe for $x$ = 0.20  (continuous curve) and $x$ = 0.40 (broken curve) for Nd$_{1-x}$Sr$_{x}$CoO$_3$. The upper and lower insets exhibits the temperature dependence of resistivity ($\rho$) for $x$ = 0.20 and 0.40, respectively.}
\label{Fig. 9}
\end{figure}

Temperature variation of resistivity ($\rho$) for the samples with $x$ = 0.20 and 0.40 are shown in the upper and lower insets of Fig. 9, respectively. The semiconducting temperature dependence of $\rho$ is observed for $x$ = 0.20, while for $x$ = 0.40 nearly linear temperature dependence of $\rho$ is observed in the paramagnetic state and a sharp decrease of $\rho$ is observed below $T_C$ at 200 K. Temperature dependence of $\rho$ for both the samples are similar to the reported results \cite{stau} except for the magnitude of $\rho$. The slightly larger values of the resistivity in the present investigations might be attributed to the particle size effect, where the particle size in the present observation ($\sim$ 150 nm) is less than that of the results reported by Stauffer {\it et al}. \cite{stau} For $x$ = 0.20 the system is close to the percolation limit, where the hole rich FM clusters are of small size, which grow and coalesce with increasing $x$. Therefore, the average size of the FM clusters for $x$ = 0.40 is larger than that of the sample with $x$ = 0.20, which is reflected in the temperature dependence of resistivity and the analysis of the $H_{cool}$ dependence of $M_{E}/M_{S}$. The size of the FM clusters is also crucial for the EB effect. The hysteresis loops measured at 5 K under identical condition with $H_{cool}$ = 20 kOe are shown in Fig. 9 for $x$ = 0.20 and 0.40. The values of $H_E$ and $M_{E}/M_{S}$ are  82 Oe and 0.55 $\times$ 10$^{-2}$ for $x$ = 0.40, while for $x$ = 0.20 the values are 500 Oe and 0.53 $\times$ 10$^{-2}$, respectively (see Table 1). In addition, the magnetization at 50 kOe and $H_C$ are 28.55 emu/g and 8.7 kOe for $x$ = 0.40, and 23.92 emu/g and 13.0 kOe for $x$ = 0.20, respectively. The values of  magnetization at 50 kOe indicate that the size and density of the FM clusters are increased for $x$ = 0.40 compared to $x$ = 0.20, which is reflected in $H_E$, where $H_E$ is much smaller for $x$ = 0.40. Similar features of EB effect was observed for the particle size dependence of EB phenomenon in the cluster-glass compound LaMn$_{0.7}$Fe$_{0.3}$O$_3$, where FM clusters are embedded in the SG matrix. \cite{thakur2} The sizes of the FM clusters were found to increase with increasing particle size, which resulted in a decrease in the EB effect. Negligible EB effect was observed for the particles with average size $\sim$ 300 nm.  The increase of the size of the FM clusters decreases the effective interface area, which might be the origin of weakening of the effective exchange coupling at the FM/FI interface in the present investigations. The explanation is consistent with the model proposed by Meiklejohn, which predicts the relation, $H_E \approx J_{ex}/$($M_{FM} \times t_{FM}$), where $J_{ex}$ is the exchange constant across the FM/AFM interface per unit area. \cite{meik2}  $M_{FM}$ and $t_{FM}$ are the magnetization and the thickness of the FM layer, respectively. The increase of $M_{FM}$ and $t_{FM}$ in the denominator of the above expression decreases the EB field. In the present investigation the average size of the FM clusters, analogous to $t_{FM}$, is increased for $x$ = 0.40 compared to $x$ = 0.20, which results in the decrease of EB field. 

\section{Summary}
As none before, we observe the signature of exchange bias effect in  Nd$_{1-x}$Sr$_{x}$CoO$_3$ for $x$ = 0.20 and 0.40 at the spontaneous  ferromagnetic/ferrimagnetic interface. When the sample was cooled in a static magnetic field, the systematic shifts of the magnetic hysteresis loops are observed as a function of temperature and cooling field. The exchange bias effect vanishes above the ferrimagnetic transition ($T_{FI}$) temperature, which indicates that the ferrimagnetic spins apply a pinning force on the reversible ferromagnetic spins at the interface below $T_{FI}$ and the pinned ferromagnetic spins give rise to the exchange bias phenomenon. The exchange bias is further confirmed by the training effect, which could be explained satisfactorily by the spin configurational relaxation model. The exchange bias field is found to increase sharply ($\leq$ 10 kOe) with increasing cooling field and then it shows a very small decreasing trend ($\geq$ 20 kOe) for high cooling field. The coercivity almost follows a similar trend of exchange bias field, where a linear dependence of exchange bias field and coercivity against cooling field is observed above 20 kOe for both the compounds. The cooling field dependence of exchange bias effect is analysed by the simplified exchange interaction model, which gives a rough estimate of the average size of the ferromagnetic clusters around $\sim$ 20 and $\sim$ 40 \AA, where the FM clusters consisting of single magnetic domain are suggested for $x$ = 0.20 and 0.40, respectively. The sizes of the ferromagnetic clusters are close to the percolation threshold for $x$ = 0.20, which grow and coalesce with increasing $x$. The large size of the ferromagnetic clusters leads to the weak exchange bias effect for $x$ = 0.40. 

%\section{Acknowledgments}
\par
One of the authors (S.G.) wishes to thank DST (Project No. SR/S2/CMP-46/2003), India for the financial support. The particle size of the samples were measured by TEM under the scheme of Nanoscience and Nanotechnology initiative of DST at IACS, Kolkata, India. The authors wish to thank Mr. A. Karmakar for the English corrections. M.P. also wishes to thank CSIR for the JRF fellowship. 

%\section{References}


\begin{thebibliography}{99}
\bibitem{meik} W. H. Meiklejohn and C. P. Bean, Phys. Rev. {\bf 102}, 1413 (1956). 
\bibitem{stamps} R. L. Stamps, J. Phys. D: Appl. Phys. {\bf 33}, R247 (2000).
\bibitem{kiwi} M. Kiwi, J. Magn. Magn. Mater. {\bf 234}, 584 (2001). 
\bibitem{nogues1} J. Nogues, J. Sort, V. Langlais, V. Skumryev, S. Surinach, J. S. Munoz, and M. D. Baro, Physics Reports {\bf 422}, 65 (2005); J. Nogues, J. Sort, V. Langlais, V.  Skumryev, S. Surinach, J. S. Munoz, and M. D. Baro, Int. J. Nanotechnology {\bf 2}, 23 (2005); J. Nogues and I. K. Schuller, J. Magn. Magn. Mater. {\bf 192}, 203 (1999). 
\bibitem{berk} A. E. Berkowitz and K. Takano, J. Magn. Magn. Mater. {\bf 200}, 552 (1999).
\bibitem{kou} J. S. Kouvel, J. Phys. Chem. Solids {\bf 21}, 57 (1961).
\bibitem{nieb} D. Niebieskikwiat and M. B. Salamon, Phys. Rev. B {\bf 72}, 174422 (2005).
\bibitem{qian} T. Qian, G. Li, T. Zhang, T. F. Zhou, X. Q. Xiang, X. W. Kang, and X. G. Lia,   Appl. Phys. Lett. {\bf 90}, 012503 (2007).
\bibitem{patra} M. Patra, K. De, S. Majumdar, and S. Giri, Eur. Phys. J. B {\bf 58}, 367 (2007).
\bibitem{kalyan} K. De, M. Patra, S. Majumdar, and S Giri, J. Phys. D: Appl. Phys. {\bf 41},  175007 (2008).
\bibitem{de1} K. De, R. Ray, R. N. Panda, S. Giri, H. Nakamura, and T. Kohara, J. Magn. Magn. Mater. {\bf 288}, 339 (2005).
\bibitem{de2} K. De, M. Patra, S. Majumdar, and S. Giri, J. Phys. D: Appl. Phys. {\bf 40},  7614 (2007).
\bibitem{liu} X. J. Liu, Z. Q. Li, A. Yu, M. L. Liu, W. R. Li, B. L. Li, P. Wu, H. L. Bai,  and E. Y. Jiang, J. Magn. Magn. Mater. {\bf 313}, 354 (2007).
\bibitem{thakur2} M. Thakur, M. Patra, K. De, S. Majumdar, and S. Giri, J. Phys. Condensed Matter. {\bf 20}, 195215 (2008).
\bibitem{luo} W. Luo and F. Wang, Appl. Phys. Lett. {\bf 90}, 162515 (2007).
\bibitem{stau} D. D. Stauffer and C. Leighton, Phys. Rev. B {\bf 70}, 214414 (2004). 
\bibitem{krim} A. Krimmel, M. Reehuis, M. Paraskevopoulos, J. Hemberger, and A. Loidl, Phys. Rev. B {\bf 64}, 224404 (2001).
\bibitem{para} M. Paraskevopoulos, J. Hemberger, A. Krimmel, and A. Loidl, Phys. Rev. B {\bf 63}, 224416 (2001).
\bibitem{yoshi} K. Yoshii, A. Nakamura, H. Abe, M. Mizumaki, and T. Muro, J. Magn. Magn. Mater. {\bf 239}, 85 (2002); K. Yoshii and H. Abe, Phys. Rev. B {\bf 67}, 094408 (2003).
\bibitem{fonda} A. Fondado, M. P. Breijo, C. Rey-Cabezudo, M. Sanchez-Andujar, J. Mira, J. Rivas, and M. A. Senaris-Rodriguez, J. Alloys Compd. {\bf 323–324}, 444 (2001).
\bibitem{ghoshray} A. Ghoshray, B. Bandyopadhyay, K. Ghoshray, V. Morchshakov, K. Barner, I. O. Troyanchuk, H. Nakamura, T. Kohara, G. Y. Liu and G. H. Rao, Phys. Rev. B {\bf 69}, 064424 (2004).
\bibitem{wu} J. Wu and C. Leighton, Phys. Rev. B {\bf 67}, 174408 (2003). 
\bibitem{tang1} Y. K. Tang, Y. Sun, and Z. H. Cheng, Phys. Rev B {\bf 73}, 174419 (2006).
\bibitem{tang2} Y. K. Tang, Y. Sun, and Z. H. Cheng, J. Appl. Phys. {\bf 100}, 023914 (2006).
\bibitem{nogues3} J. Nogu\'{e}s, C. Leighton, and Ivan K. Schuller, Phys. Rev. B {\bf 61}, 1315 (2000).
\bibitem{patra1} M. Patra, S. Majumdar, and S. Giri (unpublished). 
\bibitem{gesh} J. Geshev, J. Magn. Magn. Mater. {\bf 320}, 600 (2008).
\bibitem{nogues} G. Salazar-Alvarez, J. Sort, S. Suri\~{n}ach, M. D. Bar\'{o}, and
J. Nogu\'{e}s, J. Am. Chem. Soc. {\bf 129}, 9102 (2007).
\bibitem{binek} C. Binek, Phys. Rev. B {\bf 70}, 014421 (2004).
\bibitem{canet} F. Canet, S. Mangin, C. Bellouard, and M. Piecuch, Europhys. Lett. {\bf 52},  594 (2000).
\bibitem{mangin} S. Mangin, F. Montaigne, and A. Schuhl, Phys. Rev. B {\bf 68}, 140404(R)  (2003). 
\bibitem{berk1} A. E. Berkowitz, G. F. Rodriguez, J. I. Hong, K. An, T. Hyeon, N. Agarwal, D. J. Smith, and E. E. Fullerton, Phys. Rev. B {\bf 77}, 024403  (2008).
\bibitem{hauet} T. Hauet, J. A. Borchers, P. Mangin, Y. Henry, and S. Mangin, Phys. Rev. Lett. {\bf 96}, 067207 (2006).
\bibitem{cai} J. W. Cai, Kai Liu, and C. L. Chien, Phys. Rev. B {\bf 60}, 72 (1999).
\bibitem{bianco} L. DelBianco, D. Fiorani, A. M. Testa, E. Bonetti, and L. Signorini, Phys. Rev. B {\bf 70}, 052401 (2004); D. Fiorani, L. Del Bianco, A. M. Testa and K. N. Trohidou, J. Phys.: Condens. Matter {\bf 19}, 225007 (2007). 
\bibitem{mark} V. Markovich, I. Fita, A. Wisniewski, R. Puzniak, D. Mogilyansky, L. Titelman, L. Vradman, M. Herskowitz, and G. Gorodetsky, Phys. Rev. B {\bf 77}, 054410 (2008). 
\bibitem{meik2} W. H. Meiklejohn, J. Appl. Phys. {\bf 33}, 1328 (1962). 

\end{thebibliography}
\end{document}